# *Free-Space distribution of entanglement and single photons over 144 km*

R. Ursin[1], F. Tiefenbacher[1,2], T. Schmitt-Manderbach[3,4], H. Weier[4], T. Scheidl[1,2], M. Lindenthal[2], B. Blauensteiner[1], T. Jennewein[2], J. Perdigues[5], P. Trojek[3,4], B. Ömer[6], M. Fürst[4], M. Meyenburg[6], J. Rarity[7], Z. Sodnik[5], C. Barbieri[8], H. Weinfurter[3,4], A. Zeilinger[1,2]

[1] Institute for Experimental Physics, University of Vienna, A-1090 Vienna, Austria

[2] Institute for Quantum Optics and Quantum Information, Austrian Academy of Sciences, A-1090 Vienna, Austria

[3] Max-Planck-Institut für Quantenoptik, D-85748 Garching, Germany

[4] Department für Physik, Ludwig-Maximilians University, D-80799 Munich, Germany

[5] European Space Agency, 2200 AG Noordwijk, The Netherlands

[6] Business Unit Quantum Technology, ARC Seibersdorf Research GmbH, A-1220 Vienna, Austria

[7] Department of Electrical and Electronic Engineering, University of Bristol, Bristol, BS8 1UB, United Kingdom

[8] Department of Astronomy, University of Padova, I-35122, Italy

Contact: Rupert.Ursin@univie.ac.at

**Quantum Entanglement is the essence of quantum physics [1] and inspires fundamental questions about the principles of nature. Moreover it is also the basis for emerging technologies of quantum information processing such as quantum cryptography [2, 3], quantum teleportation [4, 5, 6, 7, 8] and quantum computation [9]. Bell's discovery [11], that correlations measured on entangled quantum systems are at variance with a local realistic picture led to a flurry of experiments [12, 13] confirming the quantum predictions. However, it is still experimentally undecided whether quantum entanglement can survive global distances, as predicted by quantum theory. Here we report the violation of the Clauser-Horne-Shimony-Holt (CHSH) inequality [25] measured by two observers separated by 144 km between the Canary Islands of La Palma and Tenerife via an optical free-space link using the Optical Ground Station (OGS) of the European Space Agency (ESA). . Furthermore we used the entangled pairs to generate a quantum cryptographic key under experimental conditions and constraints characteristic for a Space-to-ground experiment. The distance in our experiment exceeds all previous free-space experiments by more than one order of magnitude and exploits the limit for ground-based free-space communication; significantly longer distances can only be reached using air- or space-based platforms. The range achieved thereby demonstrates the feasibility of quantum communication in space, involving satellites or the International Space Station (ISS).**

Quantum theory predicts that correlations based on quantum entanglement should be maintained over arbitrary distances. Up to now, this prediction of quantum mechanics has been verified over distances of up to 13 km [14, 15, 16] using polarisation entangled photons via free-space links through the atmosphere. For time-bin entanglement a 10 km link was demonstrated in optical fibres [17] and a 50 km experiment was done in coiled fiber. In order to go well beyond all the existing tests of this prediction, it is necessary to significantly expand the distance between entangled particles. On the basis of present fibre and detector technology, it has been determined that absorptive losses and the dark counts in the detectors limit the distance for distributing entanglement to the order of 100 km [18]. One approach to overcome this limitation is the implementation of quantum repeaters which are yet not fully developed [19]. Another approach is using free-space links, involving satellites in space for bridging distances on a global scale and eventually establishing a world wide quantum communication network. The experiment described here represents a crucial step towards this goal.

A schematic layout of the experimental setup on the Canary Island is shown in Fig.1. Polarization entangled photon pairs were generated on Roque de los Muchachos (2400 m above sea level) on the island of La Palma. A picosecond-pulsed Nd:Vanadate laser emitting light at 355 nm wavelength, a repetition rate of 249 MHz and an average power of 150 mW pumped a β-barium borate crystal in a type-II scheme of spontaneous parametric down conversion (SPDC) [20]. The photons were coupled into single mode optical fibres selecting energy degenerate pairs of entangled photons with a wavelength of 710 nm with a bandwidth of 3 nm. When detecting both photons locally, we were able to observe single count rates of 1 million counts per seconds (Mcps) each, and 145000 coincidence events per second. The SPDC source produced polarisation entangled photon pairs close to the singlet state

$$\left|\psi^{-}\right\rangle = \frac{1}{\sqrt{2}}\left(\left|H\right\rangle_{A}\left|V\right\rangle_{B} - \left|V\right\rangle_{A}\left|H\right\rangle_{B}\right),$$

where $\left|H\right\rangle$ and $\left|V\right\rangle$ represent horizontally and vertically polarized photon states respectively, and the subscripts A and B label the spatial modes. In the singlet state the polarisation measurement results are (anti-) correlated in any basis. Locally, we observed that the polarisation correlations in the H/V basis had a visibility of 98%, and in the +45°/-45° basis a visibility of 96%. Thus, for the first time, a source of high quality polarization entangled photons, capable of achieving the coincidence production rate required for a space experiment could be utilized [29].

One photon from the entangled pair was measured locally (Alice). The second photon was sent via a single mode fibre to a transmitter telescope. There, the beam was guided via a 150 mm diameter lens with 400 mm focal length (f/2.7) matching the divergence of the optical fibre over a 144 km long free-space link to Bob in the Optical Ground Station (OGS) of the European Space Agency (ESA) on Tenerife, 2400m above sea level [21].

Due to various atmospheric influences such as changes of the atmospheric layering and temperature and humidity gradients, the apparent bearing of the receiver station varied on timescales of tens of seconds to minutes. Accordingly, vertical movements appeared to be more pronounced than horizontal ones (see Fig 2a). Most classical optical communication channels prevent the beam from drifting off the receiver aperture by defocusing the beam. This is not an option in single photon experiments, where maintaining the maximum link efficiency is essential. Hence in our experiment the alignment of the transmitter telescope is controlled automatically by a closed-loop tracking system using a 532 nm beacon laser shining from the OGS to the single photon transmitter.

Besides these beam drifts further processes lead to an attenuation of the optical link: beam spreading loss due to diffraction, absorption of the atmosphere and losses due to imperfections of optical components in the setup. Atmospheric losses are expected to be around 0.07 dB/km at these altitudes [22, 23]. In addition, effects due to atmospheric turbulence, such as beam wander, rapidly evolving speckle patterns and turbulence induced beam spreading, cause losses. The effective beam diameters at the OGS varied between 3.6 – 20 m depending on weather conditions. In the diffraction limited case in vacuum, the transmitter telescope would have produced a beam of 1.5 m in diameter. All these losses reduce the link efficiency but do not affect the polarization.

The OGS (Bob), a 1 m Richey-Chrétien/Coudé telescope (see Figure 1) with an effective focal length of 39 m (f/39), is used to collect the single photons with a field-of-view of 8 arcsec . The atmospheric turbulence also caused significant beam wander in the focal plane of the telescope of up to 3 mm in the worst case. Analyzing this beam wander by taking time averaged images on a CCD camera one obtains a Fried-parameter [24] of $r_0 \sim 1$ cm. To prevent the beam wandering off the detectors we re-collimate with an additional f = 400 mm lens to pass through the polarisation analyser and a 10 nm (FWHM) filter. Finally the single photons are focused with f = 50 mm lenses onto Si avalanche photo diodes. The resulting beam size and wander is now smaller than the detector's active area of 500 μm in diameter.

We measured a link optical efficiency for single photons of -25 dB under best conditions and typically -30dB (fig 2a). From this we estimate between -8 and -12 dB to be due to atmospheric loss, and between -10 and -16 dB due to the beam spreading wider than the aperture of the receiver telescope. Optical components in the OGS Coudé focus together with the output lens of the transmitter telescope account for -2 dB attenuation. Finally our detector system (including the polarisation analyser) has a quantum efficiency ~25% equivalent to a further -6 dB of loss.

From the single photons transmitted at night-time from the source to the OGS we observed 120 cps in each of our four detectors and some 50 cps collected from background photons per detector. Together with the detector dark counts 200 cps per detector, a total count rate of 1500 cps is recorded.

Each event in one of Alice's or Bob's detectors (see Fig. 1) is locally labelled with a 64-bit tag, containing the detector channel and a time tag with a timing resolution of 156 ps. The local clocks of the time tagging system are 10 MHz oscillators directly disciplined by the Global Positioning System (GPS) with a relative drift of less than $10^{-11}$ over 100 s. Furthermore, the 1 Hz GPS synchronization provides a time-reference for Alice's and Bob's time tags and the Network Timing Protocol (NTP) is used to initiate the time-tagging within 500 ms for both parties. Bob sends his time tag data to Alice via the public internet. Alice identifies the coincident events by cross-correlating both sets of time tags using software, which determines the offset (~487 µs) and drift of the two timescales. Within a coincidence window of about 1ns the average coincidence count rate is up to 20-40 cps depending on the actual atmospheric conditions.

To demonstrate quantum entanglement between measurement results on La Palma and Tenerife, we evaluated the CHSH inequality [25]. From a particular set of polarisation correlations $E(\Phi_A, \Phi_B)$ (see Tab. 1), the "Bell parameter" $S$ is determined and is bounded by 2 in any local realistic theory. However quantum mechanical predictions violate this limit with a maximum value of $S = 2\sqrt{2} \approx 2.828$. In our experiment, typically measuring over a time of 221 s with 7058 coincidence-events in total, we found S = 2.508±0.037, demonstrating the violation of the local realistic limit by more than 13 standard deviations. One of the photon is detected directly at the source at the time when the second photon is only a few meters away. Nevertheless the two local measurements were separated by 144 km and the results showed genuine quantum correlations and thus this setup can be used for quantum communication. It should be noted that according to quantum theory neither the position of the source, nor the temporal order of the measurements defines the distance to which quantum correlations can be observed. These issues, however, become relevant if one wants to test any local realistic theory by closing the corresponding loopholes; this was not the aim of the presented experiment.

To demonstrate the applicability of our setup for quantum communication, we used the quantum entanglement between our pairs to generate a quantum cryptographic key. We aligned the polarisation compensators for maximum singlet anti-correlations in the H/V and +/- bases. These settings yielded 789 coincidences within 75 s (Figure 2). The data set was used for quantum key distribution [26, 27] implemented on Alice's and Bob's computers starting from 417 bits of raw key with 20 erroneous bits which corresponds to a QBER 4.8 % ± 1 % fully understood by the various imperfection of our experimental setup (see Fig. 2). After performing online error correction (cascade) [28] and finally privacy amplification, we distilled a secure key with a length of 178 bits in total.

We demonstrated the distribution of entanglement via optical free-space links to independent receivers separated by 144 km between the two Canary Islands La Palma and Tenerife. The observed polarization correlation due to the entanglement between two remote locations violate the CHSH-Bell inequality by more than 13 standard deviations. Furthermore, we have used these correlations to establish a quantum cryptographic key between La Palma and Tenerife. We have achieved a distance between Alice and Bob by an order of magnitude larger than in all previous experiments. Within our experiment, we have overcome the attenuation expected for a downlink from a low Earth orbit (LEO) satellite. For example the minimum distance from ISS to OGS is about 400 km, whereas the atmospheric thickness is about one order of magnitude less than in our experiment, thus yielding less attenuation compared to the horizontal link here. We have used a high precision closed-loop tracking stabilizing the link for some hours in spite of atmospheric drifts. Our new source for entangled photons was able to achieve coincidence production rates and fidelities as required in future Space experiments. We demonstrated that the OGS, developed for standard optical communication to and from satellites, can be adapted for the use in quantum communication protocols. Our results demonstrate the feasibility of satellite-based distribution of quantum entanglement which is the first step to establish a worldwide network for quantum communication [29, 30].

**Acknowledgements**
The authors wish to thank Francisco Sanchez (Director IAC) and Angel Alonso (IAC), Thomas Augusteijn and the staff of the Nordic Optical Telescope (NOT), and the staff of the Telescopio Nazionale Galileo (TNG) in La Palma for their support at the trial sites. Furthermore we thank Caslav Brukner and Johannes Kofler for helpful discussions. This work was supported by ESA under the General Studies Programme (QIPS study, ESA's contract number 18805/04/NL/HE), the Austrian Science Foundation (FWF) under project number SFB1520, the A8-Quantum information Highway project of the Bavarian High-Tech Initiative, the European projects SECOQC and QAP and by the ASAP programme of the Austrian Space Agency (FFG). Additional support was provided by the European Space Agency (ESA), the Swiss National Science Foundation (SNF) and the DOC program of the Austrian Academy of Sciences.

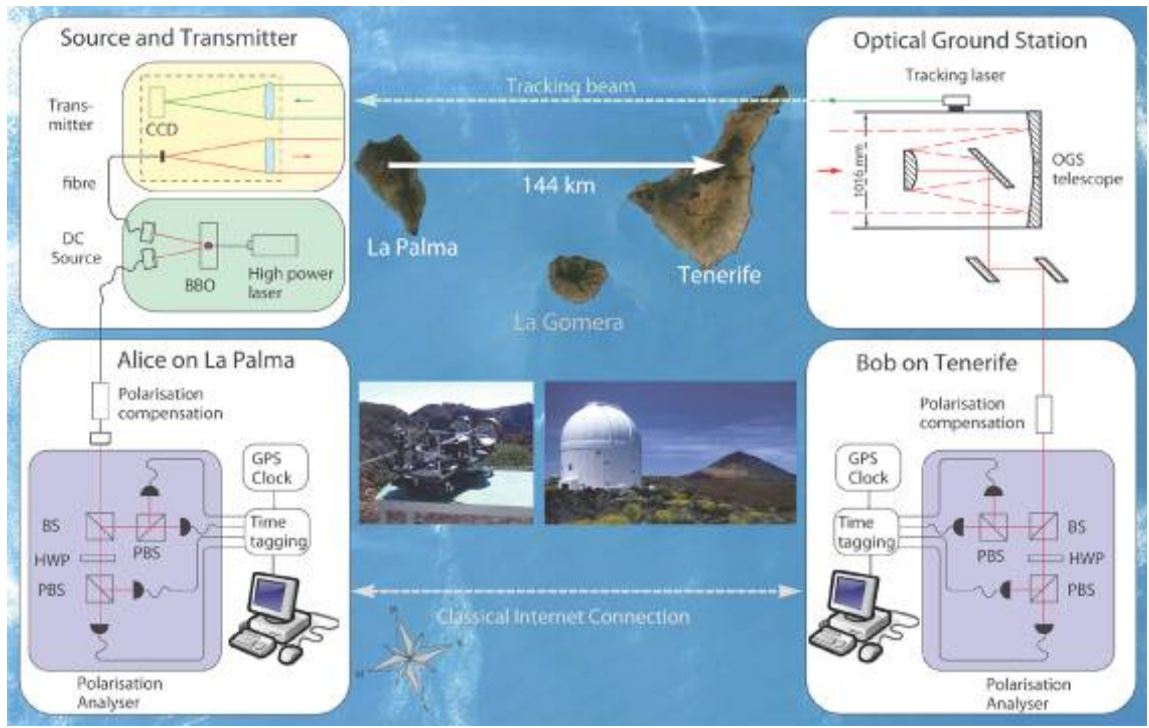

**Figure 1: The setup for free-space entanglement distribution between La Palma and Tenerife.**
**Polarisation entangled photon pairs are produced in a type-II parametric down conversion (DC)**
**source by pumping a β-barium-borate crystal (BBO) with a high power UV laser. One photon is**
**measured locally on La Palma, the other one is sent through a 15 cm transceiver lens over the 144**
**km free-space optical link to the 1 m mirror telescope of the Optical Ground Station (OGS) on the**
**island of Tenerife. The link is actively stabilised by analyzing the direction of a tracking beam (532**
**nm) sent from OGS to La Palma, which is received in a second lens focusing it on a CCD (s. Fig 2).**
**Because the tracking laser was sent in the opposite direction, no cross-talking occured to the**
**quantum channel. Both parties are using four-channel polarisation analysers, consisting of a 50/50**
**beam-splitter (BS), a half-wave plate (HWP), and two polarising beam-splitters (PBS), which**
**analyse the polarisation of an incident photon either in the H/V or in the +/-45° basis, randomly**
**split by the BS. Time-tagging units are used to record the individual times at which each detection**
**event occurs relative to a timescale disciplined by the Global Positioning System (GPS). Already**
**during data taking, Bob transmits his time tags via a public internet channel to Alice. She finds the**
**coincident photon pairs in real time by maximising the cross-correlation of these time-tags using**
**fast time-correlation software.**

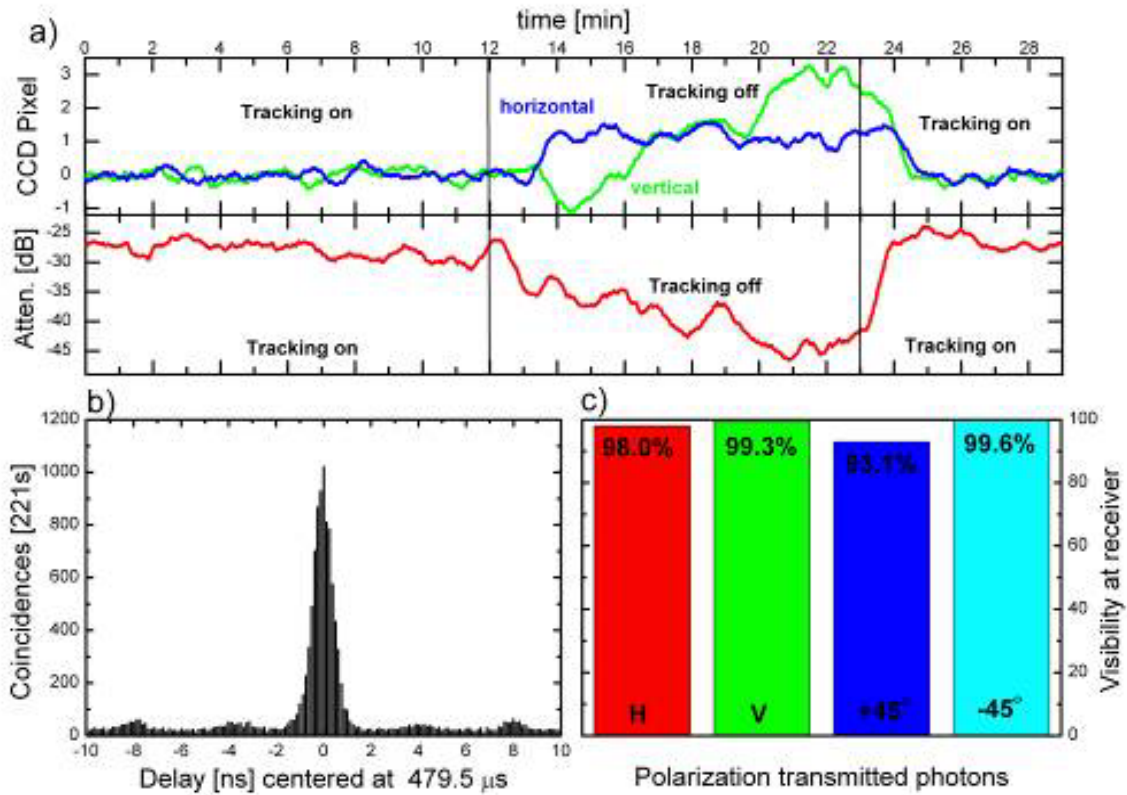

**Figure 2: Complete characterization of the quantum link:**

**(a) The Figure shows the power received at the OGS Coudé Focus from a test laser at 808 nm and the deviation of the tracking laser sampled by the CCD-camera (4.5 μm pixel size) of the transmitter telescope as a function of time. Slow changes in average pointing direction occur due to changes of the atmospheric temperature gradients and layering. To maintain maximum link efficiency over the quantum link the alignment of the transmitter platform is controlled automatically by a closed-loop tracking system. A 532 nm beacon laser sent from OGS to the transmitter is focused onto a CCD camera which is attached to the optical platform of the transmitter. The beam drifts are compensated by keeping the spot of the tracking laser on a fixed reference position by permanently readjusting the transmitter platform. Without tracking (tracking off) the beam drifts off the receiving telescope and the transmitted power decreases accordingly.**

**(b) This figure shows the distribution of occurrences of the coincidences between Alice's and Bob's detections. Centred around the flight time from Alice to Bob of about 487 μs, a clear peak occurs due to the entangled photons arriving within the coincidence window of 0,8 ns. The side peaks occurring with a period of 4ns are due to the pulsed nature of our entangled photon source (249MHz). This feature was used in our data analysis, detector clicks between the pulses were suppressed and not used, improving the fidelity of the results.**

**(c) This diagram shows the visibilities obtained using a polarized test laser beam at 808 nm wavelength transmitted over the 144 km link. The polarization is measured in the four-channel polarisation analyser in the OGS Coudé focus in a time interval of 10 min. After polarization compensation the residual visibility is shown and is constant. This shows that any polarization drifts and depolarizing effect can be neglected here.**

| $\Phi_A, \Phi_B$ | 0°, 22.5° | 0°, 67.5° | 45°, 22.5 | 45°, 67.5° |
|---|---|---|---|---|
| $E(\Phi_A, \Phi_B)$ | -0.775±0.015 | 0.486±0.020 | -0.435±0.023 | -0.812±0.014 |

**Tab. 1: Experimentally determined correlation coefficients to test for the violation of a Clauser-Horne-Shimony-Holt (CHSH)-type Bell inequality [25]. Each coefficient is inferred from the measured coincidence count rates $N_{ij}(\Phi_A; \Phi_B)$ for outcomes i and j and the settings $\Phi_A$ and $\Phi_B$, respectively.**

$$E(\Phi_A, \Phi_B) = \frac{N_{++}(\Phi_A, \Phi_B) + N_{--}(\Phi_A, \Phi_B) - N_{+-}(\Phi_A, \Phi_B) - N_{-+}(\Phi_A, \Phi_B)}{N_{++}(\Phi_A, \Phi_B) + N_{--}(\Phi_A, \Phi_B) + N_{+-}(\Phi_A, \Phi_B) + N_{-+}(\Phi_A, \Phi_B)}$$

**with i,j $\in$ {+,-}, where "+" and "-" label the outputs of a two-channel polarization analyzer. We conservatively estimated the error as the standard deviation of the Poissonian count rate distribution. The correlation coefficients are combined to yield the value of the parameter S:**

$$S = E(\Phi_A, \Phi_B) - E(\Phi_A, \Phi_B') + E(\Phi_A', \Phi_B) + E(\Phi_A', \Phi_B').$$

**Here $\{\Phi_A, \Phi_B, \Phi_A', \Phi_B'\}$ are the measurement settings of Alice and Bob. According to the CHSH inequality, any local realistic model is bound by $|S| \leq 2$, whereas quantum mechanics predicts a sinusoidal dependence for $N_{ij}(\Phi_A, \Phi_B) \propto \sin^2(\Phi_A - \Phi_B)$. The maximum of $S = 2\sqrt{2}$ is found for the following set of the angles $\{\Phi_A, \Phi_B, \Phi_A', \Phi_B'\}$ = {0°,45°,22.5°,67.5°}. Combining our experimental data, we obtain the value of $S_{EXP}$ = 2.508±0.037, thereby conclusively proving the existence of the shared quantum correlations between the Canary islands La Palma and Tenerife. The necessary statistics for this violation of 13 standard deviations were obtained by integrating over 221 s.**